\documentclass[a4paper,twoside]{article}
\newcommand{\affil}[1]{$^{\rm #1}$}
\usepackage[authoryear]{natbib}
\bibpunct{(}{)}{;}{a}{}{,}
\usepackage{graphicx}
\date{} 
\newcommand{\spr}{$s$-process}
\def\n{{\rm n}}
\newcommand{\monsoon}{{\sc monsoon}}
\newcommand{\montage}{{\sc montage}}
\newcommand{\Mo}{\rm{M}_\odot}

\newcommand{\Zo}{\rm{Z}_\odot}

\newcommand{\ud}{\mathrm{d}}

\def\tcent#1{\multicolumn{1}{c}{#1}}
\def\itp#1#2{$^{#1}{\rm #2}$}

\title{\large\bf\flushleft \montage: AGB nucleosynthesis with full $s$-process
calculations}
\author{\parbox{\textwidth}{\flushleft
\vspace{-0.5cm}
{\it R.~P.~Church\affil{A,D}, S.~Cristallo\affil{B}, J.~C.~Lattanzio\affil{A}, R.~J.~Stancliffe\affil{A}, O.~Straniero\affil{B} and R.~C.~Cannon\affil{C}}\\
\vspace{0.4cm}
{\small \affil{A}\,Centre for Stellar and Planetary Astrophysics, School of
Mathematical Sciences, Monash University, Victoria 3800, Australia}\\
{\small \affil{B}\,INAF, Osservatorio Astronomico di Collurania, 64100 Teramo,
Italy}\\
{\small \affil{C}\,Textensor Limited, 37 McDonald Road, Edinburgh, EH7 4LY,
Scotland}\\
{\small \affil{D}\,Email: ross.church@sci.monash.edu.au}}}
\begin{document}
\twocolumn[
\begin{changemargin}{.8cm}{.5cm}
\begin{minipage}{.9\textwidth}
\vspace{-1cm}
\maketitle
%
%
\small{\bf Abstract:}\\ We present \montage{}, a post-processing nucleosynthesis
code that combines a traditional network for isotopes lighter than calcium with
a rapid algorithm for calculating the \spr{} nucleosynthesis of the heavier
isotopes.  The separation of those parts of the network where only
neutron-capture and beta-decay reactions are significant provides a substantial
advantage in computational efficiency.  We present the yields for a complete set
of \spr{} isotopes for a $3\,\Mo$, $Z=0.02$ stellar model, as a demonstration of
the utility of the approach.  Future work will include a large grid of models
suitable for use in calculations of Galactic chemical evolution.

\medskip{\bf Keywords:} nucleosynthesis --- stars: AGB and post-AGB --- stars:
abundances --- methods: numerical

\medskip
\medskip
\end{minipage}
\end{changemargin}
]
\small

\section{Introduction}

The asymptotic giant branch (AGB) is the final stage of the lives of low-
and intermediate-mass stars.  Evolution during the AGB phase is
dominated by thermal pulses~\citep{1965ApJ...142..855S}.  These cyclical events
consist of brief, intense helium-burning {\it pulses} interspersed with
quiescent hydrogen-burning {\it interpulse} phases.  This complex behaviour
produces significant nucleosynthesis.  Of particular significance for the origin
of the elements heavier than iron is the \spr{}.  Neutrons produced by the
$^{13}{\rm C}(\alpha,\n)^{16}{\rm O}$ and $^{22}{\rm Ne}(\alpha,\n)^{25}{\rm
Mg}$ reactions are captured on to heavy metal isotopes, in particular
\itp{56}{Fe}.  Through a sequence of neutron captures and beta decays this
process can produce isotopes as heavy as lead and bismuth.  The AGB and \spr{}
nucleosynthesis are well covered in the literature; for example, see recent
reviews of AGB evolution by \citet{2005ARA&A..43..435H} and the \spr{} by
\citet{1999ARA&A..37..239B}, as well as references therein and other
contributions in this issue.

The recent history of \spr{} nucleosynthetic calculations is dominated by the work of
Roberto Gallino and his collaborators, and is reviewed in detail elsewhere in
this volume.  Much of this work has been motivated by the abundant data provided
in recent years by the analysis of meteoritic grains
\citep{2008PASA...25....7Z,2005sdfm.book.....L} and by spectroscopic
observations of giant stars
\citep[e.g.][]{1995ApJ...450..302L,2002ApJ...579..817A}.  Such calculations are
challenging for several reasons.  One problem is the large computational effort
required.  For stars that undergo thermal pulses the AGB is, by a large factor,
the most difficult part of the evolutionary calculations.  In order to reduce
their run-time AGB structure calculations usually only include the
reactions that generate substantial quantities of energy: the pp-chains, CNO
cycles and helium-burning reactions.  This allows a small nuclear network to be
used.  For nucleosynthetic calculations, however, a much larger network is
required.  To deal with this problem a post-processing technique is commonly
employed, where the calculation of nuclear burning and mixing is decoupled from
stellar structure and evolution calculations.  The isotopes produced and
destroyed via the \spr{} represent the majority of the nuclear network.  One
technique to deal with them is to assume an average neutron-absorption
cross-section for the heavy isotopes and, by means of a fictitious particle,
count the number of neutrons that they absorb
\citep[e.g.][]{1988ApJ...326..196B}.
This then allows those neutrons to be distributed within the
\spr{} network to calculate their nucleosynthetic effect. Alternatively a
restricted {\it s}-process network can be used, containing a sub-section of the
relevant isotopes \citep[e.g.][]{2008arXiv0809.1456K}.  A third approach has
been followed by \citet{2006NuPhA.777..311S}, who present an evolutionary model
containing a full nuclear network coupled directly to their stellar evolution
code \citep[see also][]{2009ApJ...696..797C}.

We describe here a numerical technique that allows us to compute the
nucleosynthesis of all the relevant isotopes at the same time with a reasonable
computational effort.  We split the nuclear network into two parts and solve
them separately.  The {\it lower network} comprises all elements up to and
including potassium, and some calcium and scandium isotopes.  The {\it upper
network} contains all the heavier isotopes.  The code for evaluating the lower
network is briefly discussed in Section~\ref{sect:monsoon} and that for the
upper network, along with the interface between the two, is described in
Section~\ref{sect:montage}.  We describe a sample model and give its yields in
Section~\ref{sect:yield}.

\section{\monsoon{} and the lower network}
\label{sect:monsoon}
To model the nucleosynthesis of species less massive than calcium we use
\monsoon{}, a post-processing nucleosynthesis code.  It calculates the mixing and
burning of a user-specified set of isotopes simultaneously using a standard
relaxation method.  As its input \monsoon{} takes the radius, pressure,
temperature, density, mixing length and velocity as a function of mass from
separately calculated stellar structure models.  We briefly cover the details of
\monsoon's construction here; for more information the reader is referred to
\citet{1993MNRAS.263..817C}, which describes an early version of the code, and
Church et al. (in preparation).

\monsoon{} comprises a one-dimensional, spherically-symmetric model.  Material
undergoing convection is split into two streams, one moving upwards and one
downwards.  In a given timestep material is mixed into each up-flowing cell from
the cell below, and similarly into each down-moving cell from the cell above.  A
gradient in the vertical flow causes material to flow between the two cells
in the same level.  The ratio of the cross-sectional areas of the up-flowing and
down-flowing streams can be adjusted.  However, in the model computed here we
assume that the two streams have the same cross-section.

We adopt a mesh composed of several sections, the number depending on
the evolutionary phase.  For main-sequence stars the mesh is placed at
geometrically spaced intervals in mass, increasing from the core to envelope.
When burning shells develop on the RGB and AGB they are treated with two
additional sections of mesh each, also with geometrically-spaced --- but closer
--- meshes.  Cores and inter-shell regions are treated with meshes spaced at
constant intervals in mass.  As burning shells move out through the star the
sections of mesh that represent them consume points from the region above and
release points into the region below as necessary.  Otherwise the mesh is held
steady to reduce numerical diffusion.  Some special treatment is needed to
preserve sufficient resolution in the inter-shell region on the AGB and this is
discussed in Section~\ref{sect:montage}.  It is important to note that the mesh that we adopt
does not depend on that used for the structural input.  This provides a
substantial saving in computational effort; for example, the structure code must
place many points in the outer envelope to deal with ionisation zones, whereas
very little nucleosynthesis of any interest takes place in the envelope and
hence very few points are required.  Similarly we choose the timestep
independently of that used to calculate the star's structure.

Nuclear reactions are utilised in the parameterised {\sc reaclib} form of 
\citet{2000ADNDT..75....1R}.  This utilises a seven-component fit to reaction
rates that avoids the need for interpolation in large tables.  Multiple
fits may be made in the case of complex rates, in particular those exhibiting
resonances.  For the models presented in this paper we use the recommended rates
of Illiadis (private communication).

To produce a $^{13}$C pocket and hence the required source of neutrons in
low-mass stars we follow the prescription of \citep{2008arXiv0809.1456K}.  For
each thermal pulse where third dredge-up occurs a partially-mixed zone is
artificially inserted at the bottom of the convective envelope at its point of
maximum penetration into the star.  Instead of trying to model the behaviour of
the (unknown) process causing this partial mixing, we mimic its effect by
modifying the hydrogen abundance to include an exponentially decaying profile
just below the base of the convective zone.  The capture of these protons by
$^{12}$C naturally leads to a pocket of $^{13}$C forming, which burns via the
$^{13}{\rm C}(\alpha,\n)^{16}{\rm O}$ reaction to produce the dominant neutron
source for the \spr{} in low-mass AGB stars.  We find that for a star of this
mass ($M\simeq3\,\Mo$) a reasonable quantity of $s$-processing is obtained by
choosing a zone of mass roughly one tenth that of the intershell region
\citep{2008arXiv0809.1456K}.

\section{Treatment of the \spr}
\label{sect:montage}
To treat the \spr{} we use the network of \citet{2006NuPhA.777..311S} for
isotopes from $^{44}$Ca to $^{209}$Bi.  In this part of the network the only
reactions that we consider are neutron captures, beta decays and electron
captures.  In the conditions found in AGB stars proton- and alpha-capture
reactions are not significant for these isotopes.  Decays that occur rapidly are
taken to be instantaneous.
The network is terminated by the reactions 
\[^{210}{\rm Pb}(\n,\alpha\beta\beta)^{207}{\rm Pb}\]
\[^{210}{\rm Bi}(\n,\alpha\beta)^{207}{\rm Pb}\]
\[^{210}{\rm Po}(\n,\alpha)^{207}{\rm Pb}\]
and 
\[^{210}{\rm Po}(\alpha)^{206}{\rm Pb}.\]  The alpha particles produced by
these reactions are not fed back into the lower network but are retained within
the upper network and do not react.  This is unlikely to lead to significant
errors as the \spr{} occurs within helium-rich parts of the star where the
contribution of the mechanism to the helium abundance is insignificant.

In order to increase the speed of the code such that it can produce useful
results, the burning and mixing processes in the upper network are decoupled.
A burning step takes place first, using the neutron abundances provided by the
lower network, and the resulting isotopic abundances are then mixed.  The
mixing scheme used is, as in the lower network, that of
\citet{1993MNRAS.263..817C}, and again the resulting equations are solved using
a standard relaxation method.  As a result of this simplification burning only
needs to be calculated at a single meshpoint at a time, whereas mixing is only 
calculated for a single isotope at a time.  This removes the need to invert
very large matrices and hence makes the problem tractable.  It may lead to some
small loss of accuracy, but only for reactions that occur on a similar timescale
to the mixing processes.  Such reactions are, in these models, confined to the
lower network, which is why simultaneous mixing and burning is employed there.
The neutron capture reactions that dominate the upper network are in general
much faster than convective mixing, whereas most of the $\beta$ decays are
rather slower.

For reasons of simplicity the mesh and timestep used by the upper network are
taken from the lower network.  Occasionally the upper network may require a
shorter timestep to converge: in these cases the step is broken into a set of
several intervals of equal duration.

In order to resolve the $^{13}$C pocket and resulting \spr{} reactions well it
is necessary to have good mass resolution in the intershell.  We find that
a good compromise between adequate resolution and code performance can be
achieved by inserting a large number of points, of the order of 50, when the
partially mixed zone is included at the point of maximum dredge-up.  We then
allow the code to add and remove points in the usual manner, provided that the
removal of a point does not create a composition discontinuity greater than a
factor of 1.3 between any of the lower-network abundances of the newly adjacent
points.  This is empirically found to retain sufficient resolution during the
interpulse period as the pocket burns radiatively.  At the peak of the next   
pulse the intershell convection zone reaches the remainder of the $^{13}$C
pocket: the convective motions flatten the composition profile and hence the
points are removed.

\subsection{\monsoon{} and the upper network}
The interface between \monsoon{} and the upper network is effected by means of
the species $^{44}$Ca and $^{45}$Ca.  These two isotopes are included in both
networks.  Their molar fractions, $h'$, in the lower network are given,
in terms of molar fractions $h$ of species in the upper network, by
\[h'_{^{44}\rm Ca} = \sum h_i\]
\[h'_{^{45}\rm Ca} \simeq 0\]
where the sum runs over all the species in the upper network.  In the lower
network there are two reactions that involve both species:
\[^{44}{\rm Ca}(\n,\gamma)^{45}{\rm Ca}\] 
and \[^{45}{\rm Ca}({\rm g})^{44}{\rm Ca}.\]  
The fictitious particle g is included to count the number of neutron captures
that the material in the upper network has undergone.  The rate of the second
reaction is chosen to be very fast, so that any neutron captured produces a g
particle in effect instantaneously.  This constrains the abundance of
\itp{45}{Ca} in the lower network to be very close to zero at all times,
although it could take a small but finite value in regions with a very large
neutron exposure.  The cross-section, $\sigma_{\rm ^{44}Ca,n}$, of the first
reaction is calculated as a weighted sum across all the \n-capture reactions in
the upper network:
\begin{equation}
\sigma_{\rm ^{44}Ca,n} = \sum h_i \sigma_{i,\n} / h'_{^{44}\rm Ca}
\end{equation}
where the sum once again runs over all the species in the upper network.  This
allows the number of neutrons captured in the upper network to be calculated
self-consistently, and subsequently used to calculate the changes in abundance
in the upper network.

\section{Model and yield}
\label{sect:yield}
\begin{figure}
\begin{center}
      \includegraphics[width=\columnwidth]{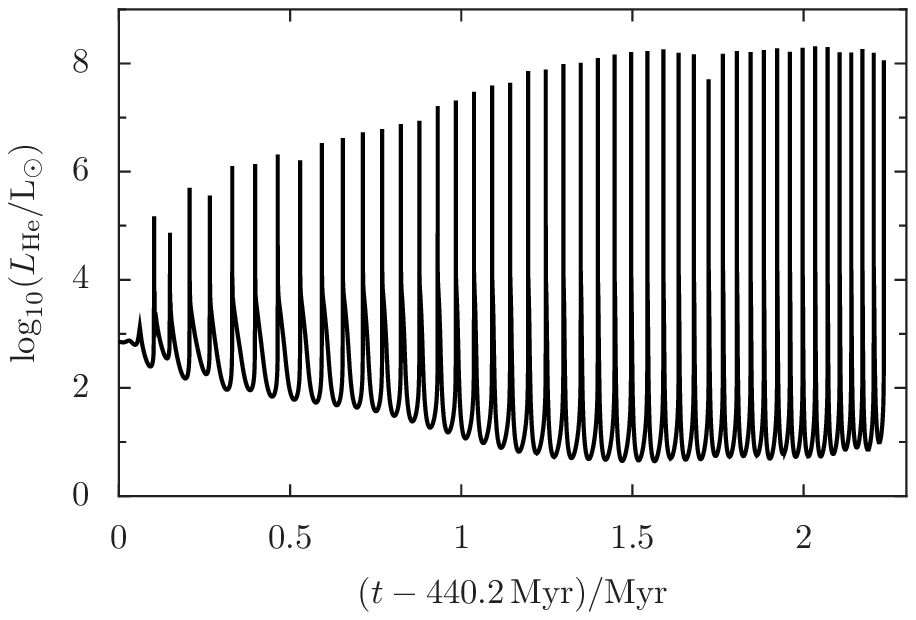}
      \caption{The helium burning luminosity of our $3\,\Mo$, $\Zo$ star is shown
      as a function of time during the AGB.  The time at the start of the plot
      is chosen to precede the onset of oscillations in the helium-burning
      luminosity.  The sharp spikes are thermal pulses, a total of 45 of which
      can be seen.}
      \label{fig:model}
   \end{center}
\end{figure}

As a sample model we present a $3\,\Mo$, $Z=0.02$ star.  The structure model
sequence was computed using the Monash version of the Mount Stromlo Stellar
Structure Programme \citep[{\sc monstar},][]{1986ApJ...311..708L,1996ApJ...473..383F}.
The opacities used by the code have been updated.  At high temperatures we use
the OPAL opacities \citep{1996ApJ...464..943I}, and at low temperatures the
opacity tables from \citet{2005ApJ...623..585F}.  We use the
\citet{1978A&A....70..227K} mass-loss law on the first giant branch and during
core helium burning with $\eta=0.4$, and the AGB mass-loss law of
\citet{1993ApJ...413..641V} thereafter.  Our initial abundances are taken from
\citet{1989GeCoA..53..197A}.  The model was evolved from the pre-main sequence
through to the late stages of the AGB where numerical instabilities prevented
any further evolution.  The helium-burning luminosity as a function of time
during the AGB is shown in Figure~\ref{fig:model}.  Dredge-up takes place during
the last 35 thermal pulses: the very final pulse does not evolve as far as dredge-up and
hence is not included.  At the end of the calculation the model has lost a total
of $1.77\,\Mo$ of material in winds and has a hydrogen-exhausted core mass of
$0.71\,\Mo$, with $0.52\,\Mo$ remaining in the envelope.  Given the mass-loss
rate of $2.16\times 10^{-5}\,\Mo\,{\rm yr^{-1}}$ and interpulse period of
$3.3\times 10^4\,{\rm yr}$ prevailing at the end of the simulation this suggests
that there are two thermal pulses missing, including the final pulse that does not
converge.

We find that we have both a larger final core mass and more thermal pulses than
the corresponding model of \citet{2007PASA...24..103K}, which was calculated
with a different version of the same code.  This is because we have used
updated physical data, including a larger mixing length parameter and updated
low-temperature opacities.  These changes lead to a later onset of the
super-wind, hence a larger core and increased number of thermal pulses.

\subsection{Nucleosynthesis during a sample pulse}
\begin{figure}
   \begin{center}
      \includegraphics[width=.75\columnwidth]{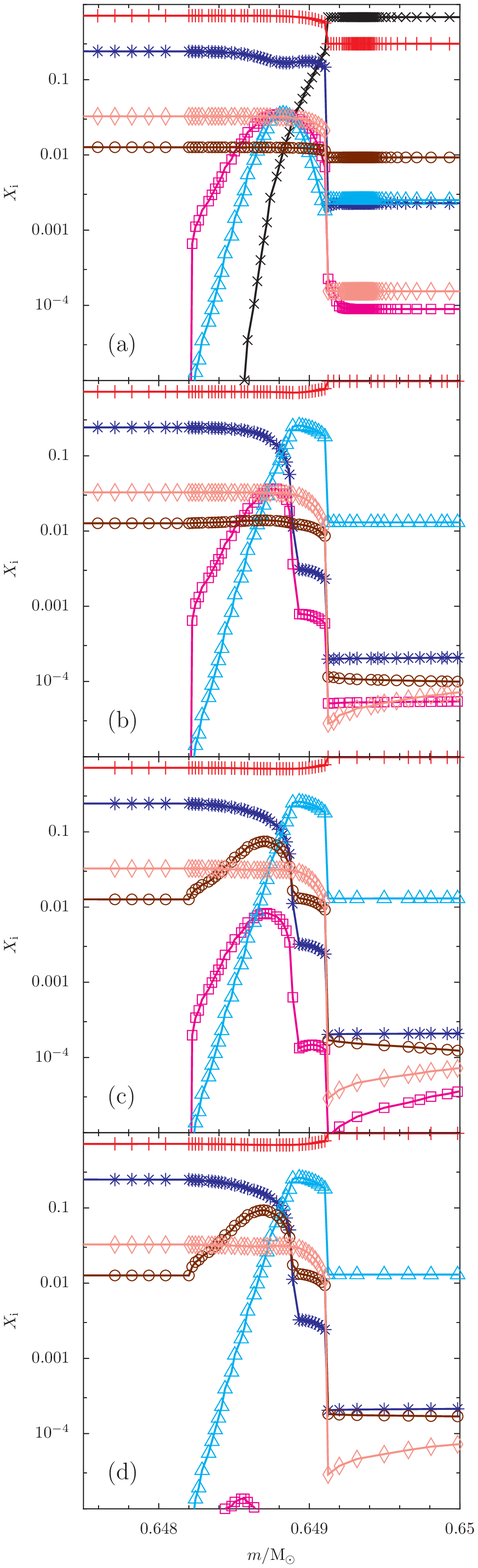}
      \caption{Isotopic abundances for structurally-significant isotopes at four
      points spanning the 12th thermal pulse and subsequent interpulse period.
      Black lines with crosses show hydrogen, red lines with plusses
      \itp{4}{He}, blue lines with stars \itp{12}{C}, magenta lines with squares
      \itp{13}{C}, cyan lines with triangles \itp{14}{N}, brown lines with
      circles \itp{16}{O} and pink lines with diamonds \itp{22}{Ne}.}
      \label{fig:pulse12structure}
   \end{center}
\end{figure}

For each pulse with dredge-up we add a partially-mixed zone as described in 
Section~\ref{sect:monsoon}.  This then goes on to form a \itp{13}{C} pocket
through the capture of protons on to \itp{12}{C}.  As an example we present the
evolution of the isotopic abundances during the twelfth thermal pulse.
Following \citet{2008arXiv0809.1456K} we add a partially-mixed zone of
mass $10^{-3}\,\Mo$ -- chosen to be about one tenth of the total intershell mass
-- at the point of maximum dredge-up.  Figure~\ref{fig:pulse12structure} shows
the effect on the most structurally-important isotopes.  

Panel (a) shows conditions $10^3\,{\rm years}$ after the maximum extent of
dredge-up.  The partially-mixed zone has been added and has partially burnt
already to produce some \itp{14}{N} and \itp{13}{C}.  Panel (b) is part-way into
the interpulse phase.  The part of the partially mixed zone with the largest
abundance of protons has burnt into CNO equilibrium, so the most abundant CNO
isotope is \itp{14}{N}.  Further in to the star a larger quantity of \itp{13}{C}
has been able to form.  Panel (c) is from about 2/3 of the way through the
interpulse period.  The majority of the \itp{13}{C} formed has burnt to
\itp{16}{O}, via the neutron-producing alpha-capture reaction.  The final panel,
(d), shows the state of the intershell just before the 13th thermal pulse.  The
\itp{13}{C} has been almost completely burned.

\begin{figure}
   \begin{center}
      \includegraphics[width=.75\columnwidth]{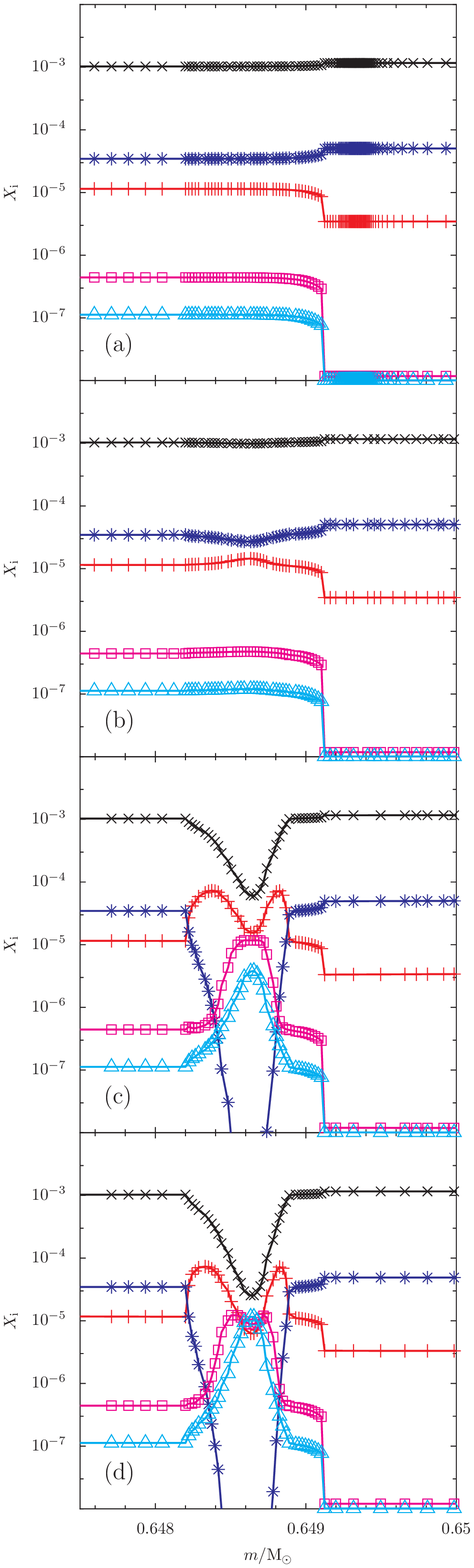}
      \caption{Isotopic abundances for selected \spr{} isotopes at the same
      times as the corresponding panels in Figure~2.  Black lines with crosses
      show \itp{56}{Fe}, red lines with plusses \itp{59}{Co}, blue lines with
      stars \itp{58}{Ni}, magenta lines with squares \itp{85}{Rb} and cyan lines
      with triangles \itp{208}{Pb}.}
      \label{fig:pulse12sproc}
   \end{center}
\end{figure}

Figure~\ref{fig:pulse12sproc} shows the effect of the \itp{13}{C} pocket and
subsequent neutron production on a selection of \spr{} isotopes.  The
discontinuity between the abundances in the envelope and those in the intershell
is caused by previous $s$-process episodes, caused by the
\itp{13}{C}$(\alpha,\n)$\itp{16}{O} reaction during the interpulse periods and
the \itp{22}{Ne}$(\alpha,\n)$\itp{25}{Mg} reaction which is eventually
activated at the pulse peaks.  The chemical signatures of these previous thermal pulses
remain within the intershell region and are mixed at each pulse peak by the
partially-overlapping intershell convection zones.  The isotopes shown were
selected to show the effect of a moderate neutron exposure on isotopes with a
range of different formation and destruction pathways and neutron-capture
cross-sections.  \itp{56}{Fe} is the initially most abundant \spr{} isotope and
hence the only discernible effect on its abundance is destruction by absorption
of neutrons.  Similarly, \itp{58}{Ni} is synthesised in supernovae, not by the
\spr: it is also only destroyed by neutron absorption.  \itp{59}{Co} is both
produced and destroyed at different mass co-ordinates in the star, depending on
the degree of neutron absorption.  Starting from the majority seed,
\itp{56}{Fe}, \itp{59}{Co} requires only three neutrons to be captured and hence
is produced in the regions of the pocket with a lower neutron exposure.  Towards
the centre of the pocket where the neutron flux is higher it is on average
depleted.  The other two isotopes, \itp{85}{Rb} and \itp{208}{Pb}, show only
production, as they lie further up the \spr{} and hence require a large number
of neutrons to be absorbed by a seed for their production.

\subsection{Comparison with single-network approach}

\begin{figure}
   \begin{center}
      \includegraphics[width=.75\columnwidth]{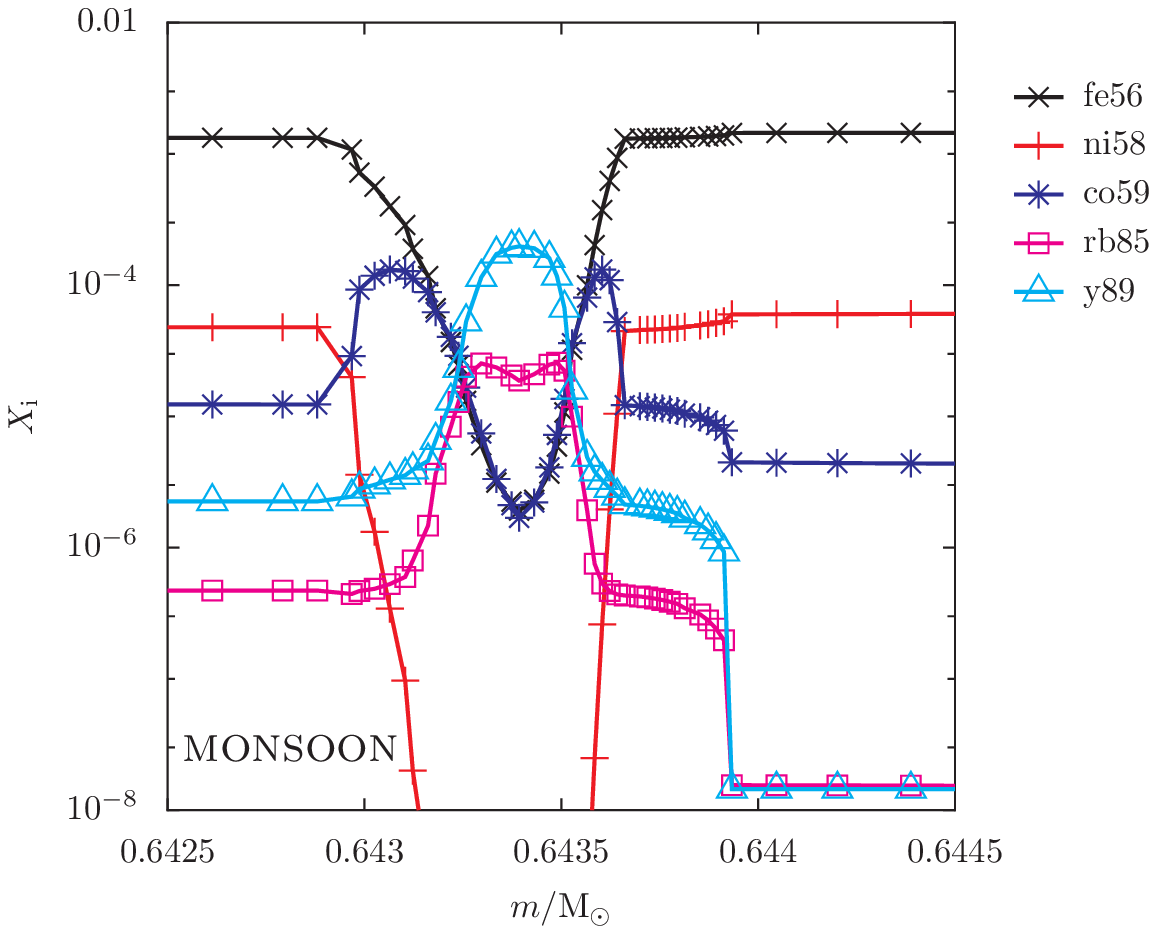}
      \includegraphics[width=.75\columnwidth]{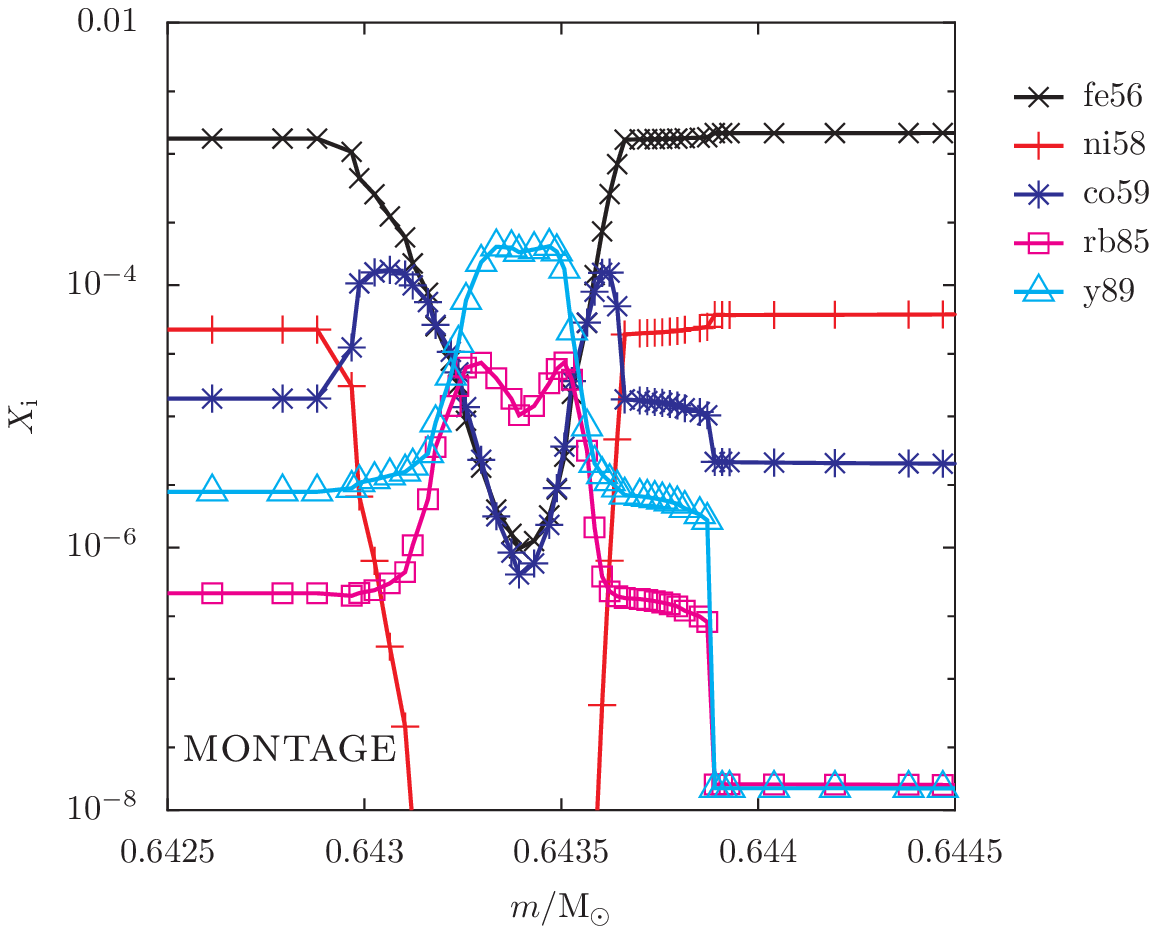}
      \caption{Isotopic abundances for selected \spr{} isotopes at the end of
      the second thermal pulse for the codes with a reduced \spr{} network.  The
      top panel shows results for \monsoon{}, the bottom panel those for
      \montage{}.  Black lines with crosses show \itp{56}{Fe}, red lines with
      plusses \itp{58}{Ni}, blue lines with stars \itp{59}{Co}, magenta lines
      with squares \itp{58}{Rb} and cyan lines with triangles \itp{89}{Y}.}
      \label{fig:compare}
   \end{center}
\end{figure}

As part of the testing process we created a version of \monsoon{} with an
extended network, that went as far as \itp{98}{Zr}.  This was used to model the
first few thermal pulses of the model, as was a version of \montage{} with a
cut-down \spr{} network.  The adopted truncation of the network reduces the
runtime for the \monsoon{} calculation to a manageable level, allowing us to
compute the models presented here in under a week.  The nuclear physics input
data for the \spr{} network in \montage{} was changed so that the same reaction
rates were used as in \monsoon{}.  A selected set of intershell abundances from
the end of the second thermal pulse in both codes are shown in
Figure~\ref{fig:compare}.  It can be seen that, phenomenologically, the
abundance profiles look very similar but that there are small differences
between the abundances calculated, particularly at the point of highest neutron
exposure.  Closer investigation of the internals of the codes revealed that this
was due to the different methods used to interpolate reaction rates in
temperature: \monsoon{} evaluates the rate for a given temperature from the {\sc
reaclib} formulae, whereas the \spr{} network in \montage{} interpolates in
$\log T$ within a pre-computed table. This leads to a systematically higher
reaction rate in \montage{} owing to the shape of the rate function at these
temperatures.  We found that the differences in abundances were consistent with
these inconsistencies in reaction rates.

\subsection{Yields}

Following~\citet{2007PASA...24..103K} we provide yields for the isotopes
that we include (Tables~\ref{tab:yields1} and \ref{tab:yields2}).  To save space
we present only the initial (Solar) value of the mass fraction for each isotope,
$X_i(0)$, and its average value in the ejecta, $\left<X_i\right>$.  Our model
still retained some envelope at the point where the evolution terminated, and
the composition of this material is included when taking the average.  Although
we would expect two additional thermal pulses to take place whilst the remaining
envelope was being lost they are unlikely to change the yield greatly as they
will represent a small fraction of the total number of thermal pulses.  A more
sophisticated approach would be to synthesise two additional thermal pulses, as
in \citet{2007PASA...24..103K}.

The yield, commonly defined as 
\begin{equation}
p_i = \int_0^\tau \left[X_i(t)-X_i(0)\right]\frac{\ud M}{\ud t}\ud t,
\end{equation}
can be found as $p_i =
\left[\left<X_i\right>-X_i(0)\right]\Delta M$, where $\Delta M=2.29\,\Mo$ is the
total mass lost by the star, including the envelope remaining at the end of the
calculation.

\begin{table*}[h]
\begin{center}
\caption{Yields from a $3\,\Mo$, $Z=0.02$ model (see text for details).
Continued in Table~2.}
\label{tab:yields1}
\footnotesize
\begin{tabular}{lrrlrrlrr}
\hline
\phantom{\small$^1_1$}Isotope $i$ & \tcent{$\left<X_i\right>$} &  \tcent{$X_i(0)$} & Isotope $i$ & \tcent{$\left<X_i\right>$} & \tcent{$X_i(0)$} &Isotope $i$ & \tcent{$\left<X_i\right>$} &  \tcent{$X_i(0)$} \\
\hline
${\rm g}$ {\phantom{\small$^1$}}& $2.999$E$-04$ & $2.977$E$-04$ &$^{55}{\rm Fe}$ & $6.302$E$-07$ & $0.000$E$+00$& $^{94}{\rm Mo}$ & $7.128$E$-10$ & $5.829$E$-10$  \\
${\rm p}$      & $6.205$E$-01$ & $6.879$E$-01$ &    $^{56}{\rm Fe}$ & $1.148$E$-03$ & $1.167$E$-03$ &  $^{95}{\rm Mo}$ & $8.523$E$-09$ & $1.014$E$-09$  \\
$^{3}{\rm He}$ & $8.963$E$-05$ & $8.544$E$-07$ &    $^{57}{\rm Fe}$ & $3.590$E$-05$ & $2.848$E$-05$ &  $^{96}{\rm Mo}$ & $1.260$E$-08$ & $1.073$E$-09$ \\
$^{4}{\rm He}$ & $3.413$E$-01$ & $2.931$E$-01$ &    $^{58}{\rm Fe}$ & $1.095$E$-05$ & $3.688$E$-06$ &  $^{97}{\rm Mo}$ & $4.655$E$-09$ & $6.210$E$-10$ \\
$^{7}{\rm Li}$ & $7.237$E$-12$ & $3.550$E$-20$ &    $^{60}{\rm Fe}$ & $1.090$E$-07$ & $0.000$E$+00$ &  $^{98}{\rm Mo}$ & $1.629$E$-08$ & $1.585$E$-09$ \\
$^{12}{\rm C}$ & $1.743$E$-02$ & $3.035$E$-03$ &    $^{59}{\rm Co}$ & $4.632$E$-06$ & $3.444$E$-06$ &  $^{100}{\rm Mo}$ & $1.286$E$-09$ & $6.456$E$-10$\\
$^{13}{\rm C}$ & $8.063$E$-05$ & $3.654$E$-05$ &    $^{60}{\rm Co}$ & $6.540$E$-12$ & $0.000$E$+00$ &  $^{97}{\rm Tc}$ & $9.579$E$-13$ & $0.000$E$+00$ \\
$^{14}{\rm C}$ & $3.640$E$-09$ & $0.000$E$+00$ &    $^{58}{\rm Ni}$ & $4.858$E$-05$ & $5.057$E$-05$ &  $^{98}{\rm Tc}$ & $1.053$E$-13$ & $0.000$E$+00$ \\
$^{14}{\rm N}$ & $2.286$E$-03$ & $1.106$E$-03$ &    $^{59}{\rm Ni}$ & $6.309$E$-07$ & $0.000$E$+00$ &  $^{99}{\rm Tc}$ & $2.032$E$-09$ & $0.000$E$+00$ \\
$^{15}{\rm N}$ & $2.053$E$-06$ & $4.367$E$-06$ &    $^{60}{\rm Ni}$ & $2.092$E$-05$ & $2.000$E$-05$ &  $^{96}{\rm Ru}$ & $2.376$E$-10$ & $2.574$E$-10$ \\
$^{16}{\rm O}$ & $8.960$E$-03$ & $9.602$E$-03$ &    $^{61}{\rm Ni}$ & $1.493$E$-06$ & $8.804$E$-07$ &  $^{98}{\rm Ru}$ & $8.579$E$-11$ & $8.948$E$-11$ \\
$^{17}{\rm O}$ & $1.906$E$-05$ & $3.891$E$-06$ &    $^{62}{\rm Ni}$ & $3.731$E$-06$ & $2.843$E$-06$ &  $^{99}{\rm Ru}$ & $6.965$E$-10$ & $6.106$E$-10$ \\
$^{18}{\rm O}$ & $1.520$E$-05$ & $2.169$E$-05$ &    $^{63}{\rm Ni}$ & $9.193$E$-08$ & $0.000$E$+00$ &  $^{100}{\rm Ru}$ & $7.636$E$-09$ & $6.119$E$-10$\\
$^{19}{\rm F}$ & $5.940$E$-06$ & $4.055$E$-07$ &    $^{64}{\rm Ni}$ & $1.662$E$-06$ & $7.438$E$-07$ &  $^{101}{\rm Ru}$ & $2.346$E$-09$ & $8.339$E$-10$\\
$^{20}{\rm Ne}$ & $1.648$E$-03$ & $1.621$E$-03$ &   $^{63}{\rm Cu}$ & $9.166$E$-07$ & $6.103$E$-07$ &  $^{102}{\rm Ru}$ & $9.467$E$-09$ & $1.565$E$-09$\\
$^{21}{\rm Ne}$ & $5.878$E$-06$ & $4.131$E$-06$ &   $^{65}{\rm Cu}$ & $6.137$E$-07$ & $2.806$E$-07$ &  $^{104}{\rm Ru}$ & $1.672$E$-09$ & $9.446$E$-10$\\
$^{22}{\rm Ne}$ & $3.954$E$-03$ & $1.303$E$-04$ &   $^{64}{\rm Zn}$ & $1.154$E$-06$ & $1.046$E$-06$ &  $^{103}{\rm Rh}$ & $2.630$E$-09$ & $9.315$E$-10$\\
$^{23}{\rm Na}$ & $1.334$E$-04$ & $3.342$E$-05$ &   $^{66}{\rm Zn}$ & $1.007$E$-06$ & $6.186$E$-07$ &  $^{104}{\rm Pd}$ & $5.413$E$-09$ & $4.211$E$-10$\\
$^{24}{\rm Mg}$ & $5.184$E$-04$ & $5.153$E$-04$ &   $^{67}{\rm Zn}$ & $1.787$E$-07$ & $9.229$E$-08$ &  $^{105}{\rm Pd}$ & $2.176$E$-09$ & $8.522$E$-10$\\
$^{25}{\rm Mg}$ & $1.070$E$-04$ & $6.773$E$-05$ &   $^{68}{\rm Zn}$ & $8.792$E$-07$ & $4.283$E$-07$ &  $^{106}{\rm Pd}$ & $7.181$E$-09$ & $1.053$E$-09$\\
$^{26}{\rm Mg}$ & $1.349$E$-04$ & $7.768$E$-05$ &   $^{70}{\rm Zn}$ & $1.448$E$-08$ & $1.458$E$-08$ &  $^{107}{\rm Pd}$ & $1.223$E$-09$ & $0.000$E$+00$\\
$^{27}{\rm Al}$ & $6.327$E$-05$ & $5.804$E$-05$&    $^{69}{\rm Ga}$ & $1.118$E$-07$ & $4.019$E$-08$ &  $^{108}{\rm Pd}$ & $8.646$E$-09$ & $1.038$E$-09$\\
$^{28}{\rm Si}$ & $6.495$E$-04$ & $6.537$E$-04$ &   $^{71}{\rm Ga}$ & $9.499$E$-08$ & $2.744$E$-08$ &  $^{110}{\rm Pd}$ & $1.074$E$-09$ & $4.515$E$-10$\\
$^{29}{\rm Si}$ & $3.690$E$-05$ & $3.429$E$-05$ &   $^{70}{\rm Ge}$ & $1.676$E$-07$ & $4.397$E$-08$ &  $^{107}{\rm Ag}$ & $6.344$E$-10$ & $6.925$E$-10$\\
$^{30}{\rm Si}$ & $2.655$E$-05$ & $2.354$E$-05$ &   $^{72}{\rm Ge}$ & $2.059$E$-07$ & $6.045$E$-08$ &  $^{109}{\rm Ag}$ & $2.584$E$-09$ & $6.554$E$-10$\\
$^{31}{\rm P}$ & $1.112$E$-05$ & $8.163$E$-06$  &   $^{73}{\rm Ge}$ & $5.901$E$-08$ & $1.745$E$-08$ &  $^{108}{\rm Cd}$ & $5.406$E$-11$ & $3.974$E$-11$\\
$^{32}{\rm S}$ & $3.929$E$-04$ & $3.962$E$-04$ &    $^{74}{\rm Ge}$ & $2.678$E$-07$ & $8.276$E$-08$ &  $^{110}{\rm Cd}$ & $6.640$E$-09$ & $5.680$E$-10$\\
$^{33}{\rm S}$ & $3.840$E$-06$ & $3.225$E$-06$ &    $^{76}{\rm Ge}$ & $1.807$E$-08$ & $1.816$E$-08$ &  $^{111}{\rm Cd}$ & $2.701$E$-09$ & $5.874$E$-10$\\
$^{34}{\rm S}$ & $1.913$E$-05$ & $1.868$E$-05$ &    $^{75}{\rm As}$ & $2.930$E$-08$ & $1.263$E$-08$ &  $^{112}{\rm Cd}$ & $8.943$E$-09$ & $1.117$E$-09$\\
$^{36}{\rm S}$ & $1.499$E$-07$ & $9.389$E$-08$ &    $^{76}{\rm Se}$ & $7.521$E$-08$ & $1.279$E$-08$ &  $^{113}{\rm Cd}$ & $2.829$E$-09$ & $5.709$E$-10$\\
$^{35}{\rm Cl}$ & $2.501$E$-06$ & $2.535$E$-06$&    $^{77}{\rm Se}$ & $3.546$E$-08$ & $1.094$E$-08$ &  $^{114}{\rm Cd}$ & $1.231$E$-08$ & $1.354$E$-09$\\
$^{36}{\rm Cl}$ & $5.508$E$-09$ & $0.000$E$+00$ &   $^{78}{\rm Se}$ & $1.268$E$-07$ & $3.442$E$-08$ &  $^{116}{\rm Cd}$ & $1.461$E$-09$ & $3.592$E$-10$\\
$^{37}{\rm Cl}$ & $1.792$E$-06$ & $8.554$E$-07$ &   $^{79}{\rm Se}$ & $2.608$E$-08$ & $0.000$E$+00$ &  $^{113}{\rm In}$ & $2.108$E$-11$ & $2.306$E$-11$\\
$^{36}{\rm Ar}$ & $7.620$E$-05$ & $7.748$E$-05$ &   $^{80}{\rm Se}$ & $2.554$E$-07$ & $7.525$E$-08$ &  $^{115}{\rm In}$ & $2.788$E$-09$ & $5.224$E$-10$\\
$^{38}{\rm Ar}$ & $1.580$E$-05$ & $1.540$E$-05$ &   $^{82}{\rm Se}$ & $1.331$E$-08$ & $1.342$E$-08$ &  $^{114}{\rm Sn}$ & $6.995$E$-11$ & $7.450$E$-11$\\
$^{39}{\rm Ar}$ & $1.962$E$-10$ & $0.000$E$+00$ &   $^{79}{\rm Br}$ & $1.790$E$-08$ & $1.243$E$-08$ &  $^{115}{\rm Sn}$ & $3.643$E$-11$ & $3.866$E$-11$\\
$^{40}{\rm Ar}$ & $6.522$E$-08$ & $2.535$E$-08$ &   $^{81}{\rm Br}$ & $3.695$E$-08$ & $1.239$E$-08$ &  $^{116}{\rm Sn}$ & $1.530$E$-08$ & $1.672$E$-09$\\
$^{39}{\rm K}$ & $3.528$E$-06$ & $3.473$E$-06$  &   $^{80}{\rm Kr}$ & $1.166$E$-08$ & $2.605$E$-09$ &  $^{117}{\rm Sn}$ & $5.088$E$-09$ & $8.902$E$-10$\\
$^{40}{\rm K}$ & $4.010$E$-08$ & $5.242$E$-09$ &    $^{81}{\rm Kr}$ & $4.839$E$-09$ & $0.000$E$+00$ &  $^{118}{\rm Sn}$ & $2.116$E$-08$ & $2.834$E$-09$\\
$^{41}{\rm K}$ & $4.194$E$-07$ & $2.634$E$-07$ &    $^{82}{\rm Kr}$ & $1.222$E$-07$ & $1.377$E$-08$ &  $^{119}{\rm Sn}$ & $6.378$E$-09$ & $1.013$E$-09$\\
$^{40}{\rm Ca}$ & $5.914$E$-05$ & $5.996$E$-05$&    $^{83}{\rm Kr}$ & $5.153$E$-08$ & $1.396$E$-08$ &  $^{120}{\rm Sn}$ & $3.009$E$-08$ & $3.879$E$-09$\\
$^{41}{\rm Ca}$ & $2.388$E$-08$ & $0.000$E$+00$ &   $^{84}{\rm Kr}$ & $2.983$E$-07$ & $7.035$E$-08$ &  $^{122}{\rm Sn}$ & $1.634$E$-09$ & $5.604$E$-10$\\
$^{42}{\rm Ca}$ & $5.076$E$-07$ & $4.200$E$-07$ &   $^{85}{\rm Kr}$ & $1.039$E$-10$ & $0.000$E$+00$ &  $^{124}{\rm Sn}$ & $7.028$E$-10$ & $7.117$E$-10$\\
$^{43}{\rm Ca}$ & $1.065$E$-07$ & $8.982$E$-08$ &   $^{86}{\rm Kr}$ & $6.699$E$-08$ & $2.197$E$-08$ &  $^{121}{\rm Sb}$ & $2.625$E$-09$ & $5.343$E$-10$\\
$^{44}{\rm Ca}$ & $1.532$E$-06$ & $1.476$E$-06$ &   $^{85}{\rm Rb}$ & $5.745$E$-08$ & $1.133$E$-08$ &  $^{123}{\rm Sb}$ & $7.322$E$-10$ & $4.038$E$-10$\\
$^{46}{\rm Ca}$ & $5.266$E$-09$ & $2.959$E$-09$ &   $^{87}{\rm Rb}$ & $2.857$E$-08$ & $4.471$E$-09$ &  $^{122}{\rm Te}$ & $3.312$E$-09$ & $3.915$E$-10$\\
$^{48}{\rm Ca}$ & $1.439$E$-07$ & $1.443$E$-07$ &   $^{86}{\rm Sr}$ & $9.243$E$-08$ & $5.098$E$-09$ &  $^{123}{\rm Te}$ & $1.138$E$-09$ & $1.367$E$-10$\\
$^{45}{\rm Sc}$ & $5.754$E$-08$ & $4.069$E$-08$ &   $^{87}{\rm Sr}$ & $6.321$E$-08$ & $3.605$E$-09$ &  $^{124}{\rm Te}$ & $6.625$E$-09$ & $7.370$E$-10$\\
$^{46}{\rm Ti}$ & $2.492$E$-07$ & $2.302$E$-07$ &   $^{88}{\rm Sr}$ & $9.072$E$-07$ & $4.342$E$-08$ &  $^{125}{\rm Te}$ & $3.028$E$-09$ & $1.108$E$-09$\\
$^{47}{\rm Ti}$ & $2.142$E$-07$ & $2.146$E$-07$ &   $^{90}{\rm Sr}$ & $2.532$E$-10$ & $0.000$E$+00$ &  $^{126}{\rm Te}$ & $1.373$E$-08$ & $2.972$E$-09$\\
$^{48}{\rm Ti}$ & $2.156$E$-06$ & $2.216$E$-06$ &   $^{89}{\rm Y}$ & $1.827$E$-07$ & $1.086$E$-08$  &  $^{128}{\rm Te}$ & $6.148$E$-09$ & $5.071$E$-09$\\
$^{49}{\rm Ti}$ & $2.356$E$-07$ & $1.686$E$-07$ &   $^{90}{\rm Y}$ & $6.435$E$-14$ & $0.000$E$+00$  &  $^{130}{\rm Te}$ & $5.387$E$-09$ & $5.514$E$-09$\\
$^{50}{\rm Ti}$ & $2.625$E$-07$ & $1.689$E$-07$ &   $^{90}{\rm Zr}$ & $1.589$E$-07$ & $1.355$E$-08$ &  $^{127}{\rm I}$ & $4.176$E$-09$ & $2.953$E$-09$ \\
$^{50}{\rm V}$ & $8.952$E$-10$ & $9.402$E$-10$  &   $^{91}{\rm Zr}$ & $4.272$E$-08$ & $2.988$E$-09$ &  $^{129}{\rm I}$ & $2.874$E$-11$ & $0.000$E$+00$ \\
$^{51}{\rm V}$ & $3.845$E$-07$ & $3.826$E$-07$ &    $^{92}{\rm Zr}$ & $6.210$E$-08$ & $4.617$E$-09$ &  $^{128}{\rm Xe}$ & $3.588$E$-09$ & $2.979$E$-10$\\
$^{50}{\rm Cr}$ & $7.274$E$-07$ & $7.643$E$-07$&    $^{93}{\rm Zr}$ & $1.867$E$-08$ & $0.000$E$+00$ &  $^{129}{\rm Xe}$ & $4.828$E$-09$ & $3.748$E$-09$\\
$^{51}{\rm Cr}$ & $6.988$E$-10$ & $0.000$E$+00$ &   $^{94}{\rm Zr}$ & $6.944$E$-08$ & $4.781$E$-09$ &  $^{130}{\rm Xe}$ & $7.566$E$-09$ & $6.010$E$-10$\\
$^{52}{\rm Cr}$ & $1.514$E$-05$ & $1.533$E$-05$ &   $^{96}{\rm Zr}$ & $4.960$E$-09$ & $7.866$E$-10$ &  $^{131}{\rm Xe}$ & $5.410$E$-09$ & $3.020$E$-09$\\
$^{53}{\rm Cr}$ & $1.753$E$-06$ & $1.771$E$-06$ &   $^{93}{\rm Nb}$ & $1.618$E$-09$ & $1.678$E$-09$ &  $^{132}{\rm Xe}$ & $1.622$E$-08$ & $3.718$E$-09$\\
$^{54}{\rm Cr}$ & $6.234$E$-07$ & $4.492$E$-07$ &   $^{94}{\rm Nb}$ & $3.326$E$-13$ & $0.000$E$+00$ &  $^{134}{\rm Xe}$ & $2.066$E$-09$ & $1.390$E$-09$\\
$^{55}{\rm Mn}$ & $1.324$E$-05$ & $1.339$E$-05$ &   $^{92}{\rm Mo}$ & $8.647$E$-10$ & $9.152$E$-10$ &  $^{136}{\rm Xe}$ & $1.146$E$-09$ & $1.148$E$-09$\\
$^{54}{\rm Fe}$ & $6.855$E$-05$ & $7.114$E$-05$ &   $^{93}{\rm Mo}$ & $5.236$E$-12$ & $0.000$E$+00$ &  \\

\hline
\end{tabular}
\end{center}
\end{table*}

\begin{table*}[h]
\begin{center}
\caption{Yields from a $3\,\Mo$, $Z=0.02$ model continued from Table~1.}
\label{tab:yields2}
\footnotesize
\begin{tabular}{lrrlrrlrr}
\hline
\phantom{\small$^1_1$}Isotope $i$ & \tcent{$\left<X_i\right>$} &  \tcent{$X_i(0)$} & Isotope $i$ & \tcent{$\left<X_i\right>$} & \tcent{$X_i(0)$} &Isotope $i$ & \tcent{$\left<X_i\right>$} &  \tcent{$X_i(0)$} \\
\hline
$^{133}{\rm Cs}$ {\phantom{\large$^1$}}& $3.064$E$-09$ & $1.289$E$-09$ & $^{157}{\rm Gd}$ & $2.898$E$-10$ & $2.172$E$-10$ & $^{184}{\rm W}$ & $7.342$E$-10$ & $1.987$E$-10$ \\
$^{135}{\rm Cs}$ & $9.940$E$-10$ & $0.000$E$+00$ & $^{158}{\rm Gd}$ & $6.980$E$-10$ & $3.470$E$-10$ & $^{186}{\rm W}$ & $4.370$E$-10$ & $1.873$E$-10$ \\
$^{137}{\rm Cs}$ & $4.150$E$-11$ & $0.000$E$+00$ & $^{160}{\rm Gd}$ & $3.439$E$-10$ & $3.093$E$-10$ & $^{185}{\rm Re}$ & $2.113$E$-10$ & $9.477$E$-11$ \\
$^{134}{\rm Ba}$ & $4.353$E$-09$ & $3.862$E$-10$ & $^{159}{\rm Tb}$ & $3.224$E$-10$ & $2.558$E$-10$ & $^{187}{\rm Re}$ & $1.881$E$-10$ & $1.603$E$-10$ \\
$^{135}{\rm Ba}$ & $2.333$E$-09$ & $1.061$E$-09$ & $^{160}{\rm Dy}$ & $1.594$E$-10$ & $3.980$E$-11$ & $^{186}{\rm Os}$ & $2.085$E$-10$ & $5.186$E$-11$ \\
$^{136}{\rm Ba}$ & $1.229$E$-08$ & $1.274$E$-09$ & $^{161}{\rm Dy}$ & $3.616$E$-10$ & $3.236$E$-10$ & $^{187}{\rm Os}$ & $1.063$E$-10$ & $5.280$E$-11$ \\
$^{137}{\rm Ba}$ & $1.045$E$-08$ & $1.835$E$-09$ & $^{162}{\rm Dy}$ & $6.797$E$-10$ & $4.393$E$-10$ & $^{188}{\rm Os}$ & $6.752$E$-10$ & $4.412$E$-10$ \\
$^{138}{\rm Ba}$ & $7.528$E$-08$ & $1.180$E$-08$ & $^{163}{\rm Dy}$ & $4.174$E$-10$ & $4.315$E$-10$ & $^{189}{\rm Os}$ & $5.666$E$-10$ & $5.370$E$-10$ \\
$^{139}{\rm La}$ & $8.706$E$-09$ & $1.656$E$-09$ & $^{164}{\rm Dy}$ & $9.496$E$-10$ & $4.915$E$-10$ & $^{190}{\rm Os}$ & $1.191$E$-09$ & $8.851$E$-10$ \\
$^{140}{\rm Ce}$ & $1.962$E$-08$ & $3.797$E$-09$ & $^{163}{\rm Ho}$ & $7.238$E$-11$ & $0.000$E$+00$ & $^{192}{\rm Os}$ & $1.385$E$-09$ & $1.389$E$-09$ \\
$^{142}{\rm Ce}$ & $7.390$E$-10$ & $4.823$E$-10$ & $^{165}{\rm Ho}$ & $4.659$E$-10$ & $3.838$E$-10$ & $^{194}{\rm Os}$ & $1.072$E$-15$ & $0.000$E$+00$ \\
$^{141}{\rm Pr}$ & $2.081$E$-09$ & $6.393$E$-10$ & $^{164}{\rm Er}$ & $5.050$E$-11$ & $1.771$E$-11$ & $^{191}{\rm Ir}$ & $1.181$E$-09$ & $1.201$E$-09$ \\
$^{142}{\rm Nd}$ & $4.341$E$-09$ & $8.555$E$-10$ & $^{166}{\rm Er}$ & $5.211$E$-10$ & $3.741$E$-10$ & $^{193}{\rm Ir}$ & $1.979$E$-09$ & $2.039$E$-09$ \\
$^{143}{\rm Nd}$ & $9.027$E$-10$ & $3.868$E$-10$ & $^{167}{\rm Er}$ & $3.257$E$-10$ & $2.571$E$-10$ & $^{192}{\rm Pt}$ & $1.630$E$-10$ & $5.273$E$-11$ \\
$^{144}{\rm Nd}$ & $2.363$E$-09$ & $7.611$E$-10$ & $^{168}{\rm Er}$ & $6.655$E$-10$ & $3.020$E$-10$ & $^{193}{\rm Pt}$ & $1.465$E$-12$ & $0.000$E$+00$ \\
$^{145}{\rm Nd}$ & $5.489$E$-10$ & $2.673$E$-10$ & $^{170}{\rm Er}$ & $3.177$E$-10$ & $1.699$E$-10$ & $^{194}{\rm Pt}$ & $2.584$E$-09$ & $2.247$E$-09$ \\
$^{146}{\rm Nd}$ & $2.006$E$-09$ & $5.573$E$-10$ & $^{169}{\rm Tm}$ & $2.712$E$-10$ & $1.716$E$-10$ & $^{195}{\rm Pt}$ & $2.334$E$-09$ & $2.321$E$-09$ \\
$^{148}{\rm Nd}$ & $3.658$E$-10$ & $1.893$E$-10$ & $^{170}{\rm Yb}$ & $1.397$E$-10$ & $3.399$E$-11$ & $^{196}{\rm Pt}$ & $2.467$E$-09$ & $1.739$E$-09$ \\
$^{150}{\rm Nd}$ & $1.773$E$-10$ & $1.879$E$-10$ & $^{171}{\rm Yb}$ & $3.077$E$-10$ & $1.603$E$-10$ & $^{198}{\rm Pt}$ & $5.222$E$-10$ & $5.013$E$-10$ \\
$^{145}{\rm Pm}$ & $2.278$E$-15$ & $0.000$E$+00$ & $^{172}{\rm Yb}$ & $5.841$E$-10$ & $2.469$E$-10$ & $^{197}{\rm Au}$ & $1.163$E$-09$ & $1.003$E$-09$ \\
$^{144}{\rm Sm}$ & $2.883$E$-11$ & $3.063$E$-11$ & $^{173}{\rm Yb}$ & $3.432$E$-10$ & $1.828$E$-10$ & $^{198}{\rm Hg}$ & $8.556$E$-10$ & $1.914$E$-10$ \\
$^{146}{\rm Sm}$ & $2.254$E$-13$ & $0.000$E$+00$ & $^{174}{\rm Yb}$ & $1.165$E$-09$ & $3.627$E$-10$ & $^{199}{\rm Hg}$ & $6.248$E$-10$ & $3.256$E$-10$ \\
$^{147}{\rm Sm}$ & $3.150$E$-10$ & $1.513$E$-10$ & $^{176}{\rm Yb}$ & $2.441$E$-10$ & $1.465$E$-10$ & $^{200}{\rm Hg}$ & $1.413$E$-09$ & $4.480$E$-10$ \\
$^{148}{\rm Sm}$ & $4.837$E$-10$ & $1.148$E$-10$ & $^{175}{\rm Lu}$ & $2.720$E$-10$ & $1.653$E$-10$ & $^{201}{\rm Hg}$ & $6.436$E$-10$ & $2.569$E$-10$ \\
$^{149}{\rm Sm}$ & $2.002$E$-10$ & $1.411$E$-10$ & $^{176}{\rm Lu}$ & $2.384$E$-11$ & $4.420$E$-12$ & $^{202}{\rm Hg}$ & $2.005$E$-09$ & $5.850$E$-10$ \\
$^{150}{\rm Sm}$ & $3.639$E$-10$ & $7.618$E$-11$ & $^{176}{\rm Hf}$ & $1.751$E$-10$ & $3.704$E$-11$ & $^{204}{\rm Hg}$ & $1.453$E$-10$ & $1.358$E$-10$ \\
$^{151}{\rm Sm}$ & $1.151$E$-11$ & $0.000$E$+00$ & $^{177}{\rm Hf}$ & $2.131$E$-10$ & $1.331$E$-10$ & $^{203}{\rm Tl}$ & $9.433$E$-10$ & $2.913$E$-10$ \\
$^{152}{\rm Sm}$ & $4.923$E$-10$ & $2.785$E$-10$ & $^{178}{\rm Hf}$ & $5.810$E$-10$ & $1.964$E$-10$ & $^{205}{\rm Tl}$ & $2.117$E$-09$ & $7.023$E$-10$ \\
$^{154}{\rm Sm}$ & $2.998$E$-10$ & $2.399$E$-10$ & $^{179}{\rm Hf}$ & $2.269$E$-10$ & $9.861$E$-11$ & $^{204}{\rm Pb}$ & $1.267$E$-09$ & $3.267$E$-10$ \\
$^{151}{\rm Eu}$ & $1.978$E$-10$ & $1.840$E$-10$ & $^{180}{\rm Hf}$ & $9.922$E$-10$ & $2.554$E$-10$ & $^{205}{\rm Pb}$ & $4.381$E$-11$ & $0.000$E$+00$ \\
$^{153}{\rm Eu}$ & $2.322$E$-10$ & $2.037$E$-10$ & $^{182}{\rm Hf}$ & $4.437$E$-11$ & $0.000$E$+00$ & $^{206}{\rm Pb}$ & $7.749$E$-09$ & $3.251$E$-09$ \\
$^{154}{\rm Eu}$ & $6.780$E$-15$ & $0.000$E$+00$ & $^{180}{\rm Ta}$ & $4.404$E$-14$ & $1.151$E$-14$ & $^{207}{\rm Pb}$ & $8.030$E$-09$ & $3.523$E$-09$ \\
$^{152}{\rm Gd}$ & $9.197$E$-12$ & $2.688$E$-12$ & $^{181}{\rm Ta}$ & $2.534$E$-10$ & $9.644$E$-11$ & $^{208}{\rm Pb}$ & $1.543$E$-08$ & $1.001$E$-08$ \\
$^{154}{\rm Gd}$ & $9.929$E$-11$ & $2.968$E$-11$ & $^{110}{\rm W}$ & $4.611$E$-13$ & $5.035$E$-13$ & $^{209}{\rm Bi}$ & $8.072$E$-10$ & $7.707$E$-10$ \\
$^{155}{\rm Gd}$ & $2.470$E$-10$ & $2.028$E$-10$ & $^{182}{\rm W}$ & $4.420$E$-10$ & $1.685$E$-10$ \\
$^{156}{\rm Gd}$ & $4.529$E$-10$ & $2.823$E$-10$ & $^{183}{\rm W}$ & $2.988$E$-10$ & $9.213$E$-11$ \\

\hline
\end{tabular}
\end{center}
\end{table*}

\section{Discussion and future work}

Our results demonstrate that we are able to efficiently calculate the
nucleosynthesis of a large number of \spr{} isotopes, incorporating both the
\itp{13}{C}$(\alpha,\n)$\itp{16}{O} and \itp{22}{Ne}$(\alpha,\n)$\itp{25}{Mg}
neutron sources.  The nucleosynthesis calculations presented here took
approximately one CPU-week of time on a modern desktop computer.  As can be seen
in Figure~\ref{fig:pulse12structure} we considerably over-resolve the intershell
region.  The number of points here could most likely be reduced, which would
lead to a substantial saving in runtime.

Our next target is to extend these results to a large grid of models.  Ideally
Galactic chemical evolution calculations need yields at masses separated by
about $0.5\,\Mo$ at intermediate masses, possibly lower at the lowest masses,
and much closer metallicity intervals than those currently available (Izzard,
private communication).  This will provide a standard,
consistent set of yields for use in calculation of Galactic chemical evolution.

\monsoon{} has hitherto always been used with the {\sc monstar} stellar evolution code.  During
testing we attempted to use the evolution code of \citet{2007MNRAS.375.1280S} to
provide output to drive \montage{}.  This worked very
well; a simple approach of writing out every tenth evolution model to the
nucleosynthesis code input file provided a frequency of data that converged
well.  Within the uncertainties provided by differences in reaction rates
between the two codes, the light element abundances predicted by the two codes
agreed well.  We conclude that \montage{} is a general post-processing
nucleosynthesis code, which with a small amount of effort we should be able to
use with {\it any} stellar evolution code, suitably modified to print out the
relevant models.  This will allow us, for example, to quantify the uncertainties
in nucleosynthesis predictions caused by different evolutionary codes in a
self-consistent manner.

\section{Summary}
{The use of a separate network and subtly different computational technique for
the main nucleosynthesis and \spr{} of a thermally-pulsing AGB star has a
substantial advantage in terms of the computational effort required.  We have
calculated the full \spr{} yields of a $3\,\Mo$ star at Solar metallicity, as a
demonstration of the utility of the approach.  We intend to extend this work to
a large grid of simulations suitable for inclusion in Galactic chemical
evolution models, as well as by making the code available as a general
post-processing tool.}


\section*{Acknowledgements}
We are very grateful to Simon Campbell for the use of his version of {\sc
monstar}.  We would also like to thank our colleagues for useful discussions, in
particular Amanda Karakas, Maria Lugaro and Carolyn Doherty.  RJS and RPC are
funded by the Australian Research Council's Discovery Projects scheme under
grants DP0879472 and DP0663447 respectively.  SC and OS are supported by the
Italian MIUR-PRIN 2006 Project `Final Phases of Stellar Evolution,
Nucleosynthesis in Supernovae, AGB stars, Planetary Nebulae'.

\bibliography{paper}


\end{document}